# Detecting and Distinguishing Majorana Zero Modes with the Scanning Tunneling Microscope

**Berthold Jäck[1,2]\*, Yonglong Xie[1,3], Ali Yazdani[1]\***

[1] Joseph Henry Laboratories & Department of Physics, Princeton University, Princeton, New Jersey 08544, USA.

[2] Present address: Department of Physics, The Hong Kong Institute of Science and Technology, Clearwater Bay, Kowloon, Hong Kong

[3] Present address: Department of Physics, Harvard University, Cambridge, Massachusetts 02318, USA

**\*Correspondence should be addressed to: bjaeck@ust.hk and yazdani@princeton.edu**

**The goal of creating topologically protected qubits using non-Abelian anyons is currently one of the most exciting areas of research in quantum condensed matter physics. Majorana zero modes (MZM), which are non-Abelian anyons predicted to emerge as localized zero energy states at the ends of one-dimensional topological superconductors, have been the focus of these efforts. In the search for experimental signatures of these novel quasi-particles in different material platforms, the scanning tunneling microscope (STM) has played a key role. The power of high-resolution STM techniques is perhaps best illustrated by their application in identifying MZM in one-dimensional chains of magnetic atoms on the surface of a superconductor. In this platform, STM spectroscopic mapping has demonstrated the localized nature of MZM zero-energy excitations at the ends of such chains, while experiments with superconducting and magnetic STM tips have been used to uniquely distinguish them from trivial edge modes. Beyond the atomic chains, STM has also uncovered signatures of MZM in two-dimensional materials and topological surface and boundary states, when they are subjected to the superconducting proximity effect. Looking ahead, future STM experiments can advance our understanding of MZM and their potential for creating topological qubits, by exploring avenues to demonstrate their non-Abelian statistics.**

1. **Introduction**

Majorana zero modes (MZM) are non-Abelian anyons that emerge as localized zero-dimensional end states of one-dimensional (1D) topological superconductors[1]. Unlike fermions and bosons, anyons are quasi-particles, whose particle interchange modifies the quantum-mechanical ground state of the host system. The ground state of a system with non-Abelian properties has multiple degenerate configurations, which are not specified by the spatial locations of the MZM. Adiabatic "braiding" of MZM provides the means to perform qubit operations within the subspace of the degenerate quantum state manifold, while their "fusion" can be used as means of qubit read out. In a qubit based on MZM, the quantum information is stored non-locally and protected by a topological energy gap of the system, and therefore it would be more resilient to local perturbation that can cause quantum decoherence[2–5]. To date, strong evidence for the existence of MZM has come from material platforms in which the proximity effect from a conventional superconductor is used in concert with strong spin-orbit and Zeeman or ferromagnetic exchange interactions to engineer a system with a topological superconducting ground state[6–20]. In particular, experiments on semiconducting nanowires[21–23] and chains of magnetic atoms[24–26] provided evidence for the existence of MZM in a condensed matter setting over the past years.

STM experiments combine the ability to visualize the atomic structure of condensed matter systems with the capability of studying their electronic structure with high energy and spatial resolution. In past two decades, these capacities have made STM a tool of choice for the investigation of quantum phases of matter at microscopic length scales[27]. Especially in the context of topological superconductivity, STM has emerged as a valuable tool to visualize the presence of MZM, which appear as localized zero-bias peaks (ZBP) in spectroscopic measurements, across a variety of topological quantum materials[24,28–32].

Despite the tremendous progress and achievements in research on topological superconductivity, skepticism of the research community toward the interpretation of localized ZBP as signatures of MZM exists. The debate on possible trivial origins of the observed ZBP[33–40] and lack of consistency of experimental results in some studies[41–43] emphasizes the key question of how trivial ZBP can be distinguished from topological ZBP in suitable experimental setups.

In this article, we will review the current state of research on topological superconductivity and MZM using STM experiments. We describe the pivotal role of STM in exploration of topological superconductivity across various material platforms, and we discuss how STM experiments with functional tips can probe other MZM properties, such as its particle-hole symmetry and spin, which can be used as a diagnostic tool to unequivocally distinguish topological from trivial ZBP[44–47]. We close by outlining existing theoretical proposals[48] and future experiments, which aim at demonstrating the non-Abelian exchange statistics and anyonic ground state degeneracy on the atomic scale using chains of magnetic atoms. Considering the recent developments in assembling such chains atom by atom with the STM tip[30,49–51], the realization of these concepts is in sight and would constitute milestone achievements in the quest toward topologically protected quantum computation.

## 2. Majorana modes in 1D chains

### 2.1 Kitaev model for topological superconductivity in spinless 1D systems

The idea of a 1D topological superconductor hosting MZM was first introduced in a simple and elegant model proposed by Kitaev in 2001[1]. It describes spinless electrons hopping on a 1D lattice in the presence of a nearest-neighbor p-wave superconducting pairing interaction. Depending on the model parameters, the system exhibits two distinct ground states (Box 1), which can be best understood by decomposing the electronic states into pairs of MZM. In the trivial regime (Box 1, upper panel), the system simply consists of ordinary fermions, i.e. pairs of MZM, located on each lattice site. In the topological regime (Box 1, center panel), pairing between the MZM of neighboring sites results in an unpaired localized MZM (red sphere) at the opposite ends of the chain. The emergence of these edge excitations at zero energy cost represents a distinct localized signature of MZM that can be detected with the STM (Box 1, lower panel). In these experiments, this signature would appear as a peak at zero bias voltage, the so-called ZBP, in the measured differential tunnel conductance (dI/dV)-spectrum. We note that the bias-voltage-dependent dI/dV-spectrum is proportional to the energy-dependent local density of states of the sample, where energy maps to bias voltage as $E=eV$.

The main challenge for realizing the Kitaev model is to make electrons behave as spinless fermions and to induce superconducting pairing among them. A fully spin-polarized system that breaks time-reversal symmetry (TRS) can be considered as having spinless electrons; however, intrinsic pairing instability between such spin-polarized electrons, so-called p-wave pairing has been elusive in nature. Various concepts have been proposed to engineer a 1D superconductor with an effective p-wave pairing[6,8–12,14–16,18–20] via the proximity effect from a conventional s-wave superconductor. The central idea is built on the combination of Rashba spin-orbit coupling (SOC) with time-reversal symmetry breaking, such as induced by an external magnetic field or ferromagnetism, to lift the spin degeneracy at the Kramer's point.

**Box 1**

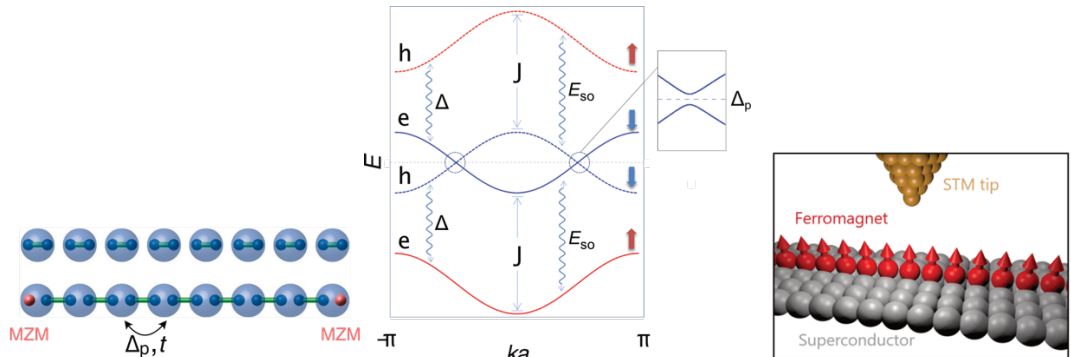

**The Kitaev model**[1] describes the hopping of spinless electrons with strength, $t$, between sites of a one-dimensional (1D) chain in the presence of p-wave superconducting pairing, $\Delta_p$. Fractionalizing the fermionic modes into pairs of Majorana quasiparticles, the system can assume topologically distinct ground states: A topologically trivial state with on-site pairing and a topologically non-trivial state with nearest-neighbor pairing of Majorana quasiparticles, respectively. Intuitively, in the topologically non-trivial case two individual Majorana quasiparticles remain localized to the chain end.

**Material realization** of Kitaev model of 1D topological superconductivity can be achieved by considering the normal state band structure of a 1D ferromagnet under the influence of Rashba SOC with amplitude $E_{SO}$[6,44]. Ferromagnetic exchange splitting of strength $J$ separates minority and majority bands. When the Fermi level lies in the minority band, only one spin species is populated, and the system can be regarded as spinless. Rashba SOC imprints a momentum-dependent spin texture on these 1D states, which facilitates proximity-induced pairing by an s-wave superconductor of the electrons in the otherwise fully spin-polarized minority band. This combination of magnetism and superconductivity, therefore, realizes a 1D topological superconducting state with p-wave pairing symmetry that can host zero-dimensional MZM at its ends.

**STM experiments** are ideally suited to study topological superconductivity and MZM across nanoscopic material platforms. In addition to the study of topographic sample properties[52], STM experiments can detect localized MZM at the ends of one-dimensional topological superconductors, such as ferromagnetic chains of atoms placed on top of a superconducting surface[24]. MZM would appear as zero-energy states inside a superconducting gap in the measured dI/dV-spectrum of scanning tunneling spectroscopy measurements. The use of functional tips diversifies the spectroscopic toolbox of STM measurements, facilitating the measurement of electron-hole symmetry[45] or spin signature of MZM[28,44,46], see Sec. 4.

One approach to realize this concept is shown in in Box 1, which depicts the normal state band structure of a 1D ferromagnet under the influence of Rashba SOC[6,44]. Due to the ferromagnetic exchange splitting, the minority and the majority bands are energetically separated. When the Fermi level lies in the minority band, only one spin species is populated, and the system can be regarded as spinless. Rashba SOC imprints a momentum-dependent spin texture on these 1D states, which facilitates proximity-induced pairing by an s-wave superconductor of the electrons in the otherwise fully spin-polarized minority band. This combination of magnetism and superconductivity, therefore, realizes a 1D topological superconducting state with p-wave pairing symmetry that can host zero-dimensional MZM at its ends.

A related approach utilizing interplay between magnetism and superconductivity that arrives at the Kitaev model considers the in-gap Shiba states[53,54] induced by magnetic atoms in a superconductor. In the limit that that overlap between the magnetic atoms is weak (unlike the band picture described above), considering only the overlap between the Shiba state, theoretical analysis shows that a topological superconductor with MZM can be created, provided there is helical order in the 1D magnetic chain[6,9–12]. The stability of such helical order in this limit has been subject of considerable theoretical studies (e.g. Ref.[55]); however, the combination of ferromagnetism with strong SOC is found to be equivalent to helical magnetic order[56], and have proven to be a more feasible pathway for realizing topological p-wave superconductivity.

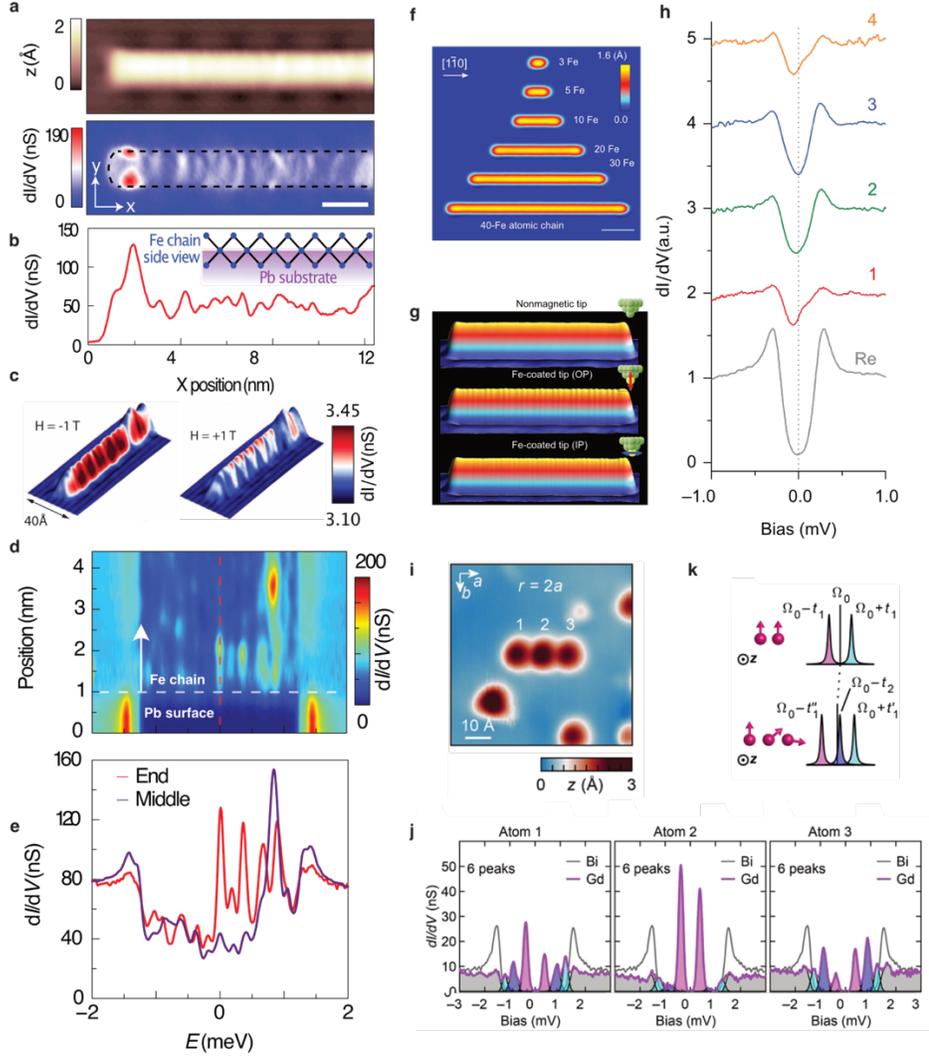

**Figure 1 | Kitaev model and atomic chain platform. a,**. STM topography (top) and zero-bias conductance map (bottom) of a Fe chain on Pb(110) substrate. **b,** Zero bias conductance profile taken along the Fe chain in panel a (see also red dashed line in panel d). The inset illustrates the conjectured zig-zag structure of Fe chain. **c,** STM topography of a Fe chain, which is overlaid with the dI/dV-signal recorded at $V_B$=30 mV with tips of opposite spin polarization, reproduced from[24]. **d,** Spectroscopic line-cut along the Fe chain axis as a function of energy and tip-position. **e,** dI/dV spectra recorded at the chain end (red) and its average in the middle of the chain (blue). **f,** Stacked STM images for the artificially constructed Fe chains of various lengths. **g**, Top: 3D-rendered STM image of a 40-atom-long Fe chain measured with a nonmagnetic PtIr tip. Spin-polarized STM images recorded with Fe-coated PtIr tips sensitive to the out-of-plane (middle panel) and the in-plane (lower panel) component of the spins in the chain. The magnetization directions of the tips are schematically depicted in the inset. **h**, dI/dV spectra obtained at the positions from rhenium surface to the bulk of the chain. Red curve, labelled 1, is taken at the chain end. **i,** Atomic assembly with the STM tip was used to build small chains of gadolinium (Gd) atoms on the surface of proximitized Bi(110). **j,** High-resolution STS measurements at milli-Kelvin temperatures of the superconducting gap on top of the Gd atoms reveal a rich dI/dV-spectrum. Six Shiba states with various spectral weight on top of different atoms are observed. **k,** These characteristics can be understood by considering the interplay of RKKY-interaction, spin-orbit coupling and surface magnetic anisotropy.

Panels a-c, d, e are reproduced from[45], panel c from[24], panels f-h from[30], and panels i-k from[51].

## 2.2 Ferromagnetic iron chains on lead

Chains of magnetically coupled atoms on the surface of a superconductor (Box 1 and Fig.1) have been used to realize the 1D Kitaev model and have been established as an novel platform to study topological superconductivity and to directly visualize the presence of MZM with STM[24–26,45,46]. In these experiments, localized ZBP within the superconducting gap were observed at the chain ends and were interpreted as the spectroscopic signature of MZM[24]. Fig. 1a shows an STM topography of an iron (Fe) chain on a lead (Pb) (110) substrate. The regularly shaped Fe chain was realized through self-assembly of evaporated Fe atoms that arrange in a linear zig-zag structure, which is illustrated in the inset of Fig. 1b. Spin-polarized STM experiments were used to demonstrate the ferromagnetic nature of the Fe chains, as shown in Fig.1c, as well as the presence of Rashba SOC on the surface of Pb. These measurements, in combination with spectroscopic measurements of the normal states of these Fe chains, and their comparison with theoretical calculation provides evidence that these chains are in a regime for topological superconductivity to emerge per the mechanism depicted in Box 1.

The existence of topological superconductivity along the chain and localized MZM at the Fe chain ends was demonstrated through high-resolution scanning tunneling spectroscopy (STS) measurements of the superconducting gap region, as shown in Fig. 1d. While the superconducting dI/dV-spectrum at the chain's bulk is dominated by energy-symmetric Shiba states[53,54], the spectrum at the chain's end features a localized peak at zero applied bias. This is also highlighted by individual point spectra in Fig. 1e and the zero-bias conductance profile in Fig. 1b. The ZBP at the chain end has no detectable splitting at milli-Kelvin temperatures, and it can be interpreted as the probability density of a MZM localized to the chain end[6,9].

Spectroscopic imaging with the STM revealed the spatial distribution of the ZBP, which shows the importance of chain-substrate hybridization in understanding such measurements. The dI/dV-map in Fig. 1a (lower panel), which was recorded at zero applied bias voltage, shows a zero-energy state at the chain end. It features an intriguing spatial pattern, whose maximum is situated near the chain sides. The comparison with model calculations, capturing this "eye" shape feature, demonstrates that substantial spectral weight of the ZBP resides in the host superconductor that is the surface of Pb(110) substrate. This finding highlights the significant role of hybridization between electronic states of chain and substrate, which is also reflected in the ZBP localization length. The observed localization length of the zero-energy state in Fig.1f (~10Å) is much shorter than the superconducting coherence length, which can be understood in terms of a strong Fermi velocity renormalization induced by chain-substrate hybridization[57]. These conclusions were further supported by experiments on Fe chains that where covered with a monolayer of Pb; STM measurements revealed the clear presence of MZM signature in this overlayer[45].

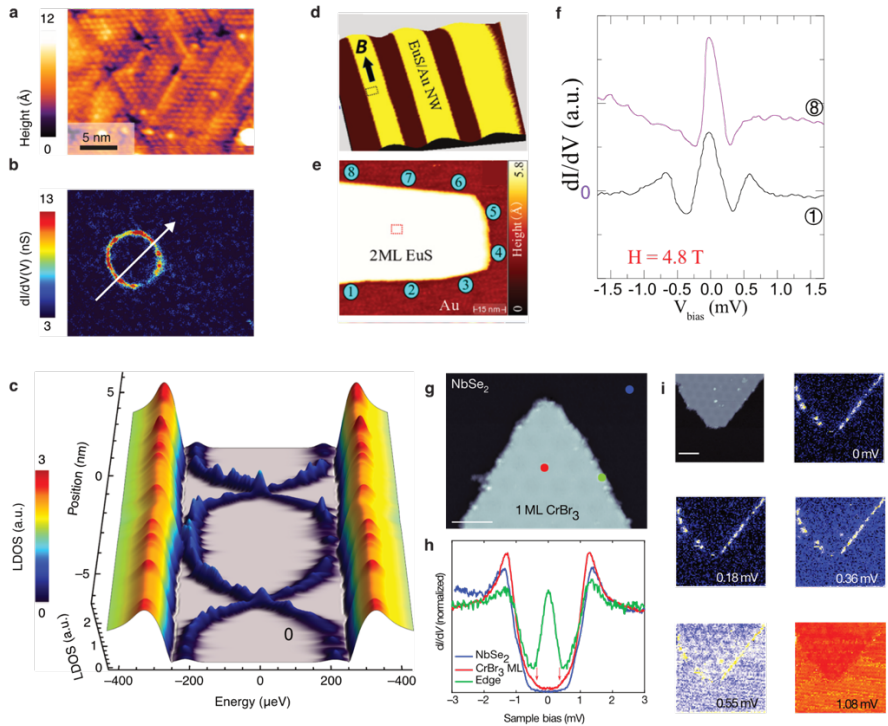

**Figure 2 | Majorana quasiparticles in 2D systems with ferromagnetism. a**, STM topography of the Pb/Co/Si(111). **b**, dI/dV map of the same area at measured at a bias voltage set point of 1.30 meV. **c**, Line-cut of the deconvoluted LDOS along the white arrow in panel b, showing the spatial dispersion of the in-gap state. The edge states display a X-shape at the interface of the cluster. **d**, Large scale STM topography of Au(111) nanowire array. **e**, STM topography of the black rectangle region in a, showing EuS island deposited on top of a gold nanowire. **f**, ZBP emerge in the dI/dV-spectra, which were measured at specific edge positions, 1 and 8, of the EuS island as indicated in panel e. **g**, STM topography of a CrBr3 monolayer deposited on the surface of NbSe$_2$. **h**, dI/dV spectra recorded with the STM tip located at different spatial positions, which are indicated by the correspondingly colored dots in panel g. **i**, Top row, left panel: STM topography of the sample, All the other panels show dI/dV maps recorded at different indicated energies. Panels a-c reproduced from[29], panels d-f reproduced from[32], and panels g-i reproduced from.[58]

### 2.3 Other Atomic Chains

A crucial figure of merit for systems hosting MZM is the size of their p-wave gap, which is $\sim \Delta E_{so}/J$, with $\Delta E_{so}$ as the spin-orbit splitting and $J$ as the ferromagnetic exchange interaction in atomic chains[6]. The topological gap protects the MZM from poisoning through quasiparticles in its vicinity, and it is key to the application of MZM for topologically protected quantum computation. Experimentally, one practical way to estimate a lower bound of the p-wave gap size is to consider the energy of the peak in a tunneling spectrum, which appears at the smallest energy above zero bias. In the case of the Fe chain platform, this energy is found to be at a modest value of around 150μeV for a peak in the chain center (corresponding to T<2 K)[24]. Comparably small gap values are also found for the semiconducting nano-wire platform[21,23], putting the MZM at risk of being poisoned through thermally excited quasiparticles[59]. Hence, a strong motivation exists for discovering other material platforms beyond the Fe chain platform with the potential to stabilize the larger p-wave gaps.

Following the approach of synthesizing Fe chains on Pb(110)[24], Ruby *et al*., prepared cobalt (Co) chains on Pb(110)[60]. However, in stark contrast to the Fe chains, the zero-energy mode appears delocalized along the Co chain and no ZBP was found at the chain ends, possibly owing to an even number of bands crossing Fermi energy. The contrasting experimental outcomes on the Fe and Co chain platforms[24,60], respectively raise the question about the best strategy to engineer 1D topological superconductors. Guidance may come from previous model calculations[6,24], which outline that an increased number of magnetic atoms in the unit cell results in a larger number of energy bands and can reduce the size of the topological phase space. It is, therefore, desirable to tailor atomic chains with preferably simple lattice structures that admit the largest topological phase space.

Kim *et al*., explored this direction and used atomic manipulation with the STM tip to manually assemble linear chains of Fe atoms on a rhenium substrate (Fig. 1f)[30]. Spin-polarized measurements with different tip magnetization directions reveal the spin-spiral arrangement of these Fe atoms (Fig. 1g), which is suitable to induce topological superconductivity as outlined in an earlier proposal[9]. Spatial dependent STS measurements on these chains display an enhanced zero bias LDOS at the chain end (Fig. 1h), when the number of atoms in the chain exceeds twelve, consistent with a topological superconducting phase and MZM. However, the small superconducting gap of Re, $\Delta_{Re}$=0.28 meV, renders the clear detection of a distinct ZBP difficult, therefore, making the presence of MZM in this system debatable.

More recent efforts have focused on extending such atomic manipulation experiments to materials, which combine a larger superconducting gap with a longer Fermi wavelength and strong spin-orbit coupling. Such system can be obtained by the epitaxial growth of bismuth(110) thin films on the surface of niobium crystal[51]. In this platform, depending on the separation of spins, the interplay between Ruderman-Kittel-Kasuya-Yosida (RKKY) interaction, spin-orbit coupling, and surface magnetic anisotropy stabilizes different types of spin alignments. These alignments influence the hybridization of the in-gap Shiba states (Fig. 1i-k) and show promise of engineering the band structure of such states for creating topological phases. They also show that spin-spin interaction can be tuned on length scale longer than interatomic distances in the presence of large proximity gap from the Nb substrate, therefore providing the possibility to create a helical spin chains atom-by-atom in the presence of robust superconductivity.

3. Majorana modes in 2D systems
3.1 From 1D chains to 2D Islands.

The idea of combining magnetism and superconductivity with strong Rashba SOC to engineer topological superconductivity can naturally be extended to two dimensions[29,58,61]. A 2D topological superconductor is predicted to harbor propagating Majorana edge modes along its one-dimensional boundary[62,63]. Such chiral 1D Majorana mode is characterized by a linear, i.e. massless, quasiparticle energy-momentum dispersion, which connects the edges of the topological superconducting gap. It is expected to give rise to a flat dI/dV-signal inside the superconducting gap, by which it can be detected in STS measurements.

The first report of realizing such a system is that of a monolayer of superconducting Pb covering Co islands on a silicon (111) substrate[29]. Fig. 2a shows an STM image of a Co island covered with Pb film. While the presence of Pb film hinders the structural characterization of the underneath Co islands, the locations of which can yet be identified by inspecting its influence on the superconducting electronic states of Pb. A Conductance map taken at zero bias inside the superconducting gap of the monolayer Pb film on the same area reveals a concentric ring-like pattern of high dI/dV amplitude, see Fig. 2b. The dispersive character of this 1D edge mode is illustrated in Fig. 2c, where the authors have examined its spatial evolution across the center of the Co island (Fig. 2b). The observation of dispersive and gapless edge modes on this and a similar sample platform[61] suggests the realization of a two-dimensional topological superconductor using arrays of magnetic atoms, potentially hosting chiral MZM.

2D islands of the magnetic insulator EuS, which are deposited on top of gold (Au) nano-ribbons, were also explored as an alternative way to engineer topological superconductivity in the Au(111) surface state (Fig. 2d,e)[18,32]. This concept is motivated by the large SOC of the Au(111) surface state, $E_{SO}$=110 meV[64], which could favor the realization of very large topological gaps. In the presence of superconductivity induced by the vanadium substrate underneath the gold nano-ribbons, the authors report the sighting of localized ZBP at the islands edges in STS measurements, when an in-plane magnetic field is applied, Fig. 2f[32]. The appearance of ZBP at specified island edge positions (1 and 8 in Fig. 3d) is, furthermore, linked to the magnetic field direction. Unlike other 2D platforms hosting 1D chiral edge modes, here theoretical modeling suggests that the directional in-plane magnetic field in concert with a magnetic exchange field underneath the EuS island could stabilize zero-dimensional MZM instead.

In addition to these material systems, van-der-Waal (vdW) heterostructures fabricated from transition-metal dichalcogenides are other promising candidates for realizing 2D p+ip topological superconductivity. Platforms based on vdW-heterostructures are insofar desirable, as they permit the controlled assembly of designer quantum materials[65,66], by stacking layers of vdW-coupled 2D materials with various properties. A recent study followed this approach and deposited monolayer islands of the magnetic insulator $CrBr_3$[67] on the surface of superconducting bulk $NbSe_2$, see Fig. 2g. Low-temperature STM experiments found a ZBP localized to the $CrBr_3$ island edge[58], see Fig. 2h, and spectroscopic imaging with the STM revealed the 1D character of this ZBP, see Fig. 2i. Based on model calculations, the authors interpret this edge state as signature of a chiral Majorana mode, and they argue that the ferromagnetic $CrBr_3$ layer induces a 2D topological superconducting phase with Chern number three in the top-most layer of the bulk $NbSe_2$ substrate. However, 1D chiral Majorana modes with approximate linear dispersion around zero energy are expected to produce a flat but not a peaked LDOS inside the superconducting gap[68]. Further experiments will, therefore, be necessary to address the deviating spectral characteristics of the edge state LDOS in this study and to elucidate its inhomogeneous spatial pattern, see Fig. 2h, which contrasts the results from STM experiments on other candidate material platforms[29,61,68].

**3.2 Majorana vortex core states in two-dimensional topological superconductors.**

One- and two-dimensional topological boundary states of TRS topological insulators[69] provide alternative pathways to realize MZMs in condensed matter systems[70]. Owing to the non-trivial bulk topology, their Fermion doubling is lifted. These states, therefore, provide a natural platform for realizing TRS topological superconductivity via the proximity effect from adjacent bulk s-wave superconductors[13,71–73]. The helicity of the topological boundary modes favors the realization of large topological s-wave gaps with sizes comparable to the gap size of the host superconductors on the order of milli-electron volts. This presents a clear advantage over the discussed p-wave superconductors, where the gap size is reliant on the strength of the spin-orbit coupling[74] and may be far smaller than s-wave superconductor used for the proximity effect[21,24]. However, the challenges to realize topological superconductivity in material platforms[75–78], which host TRS topological boundary modes, lies in isolating the effect of superconductivity on bulk of such materials as compared to their boundary modes, where we expect superconductivity to have a topological nature.

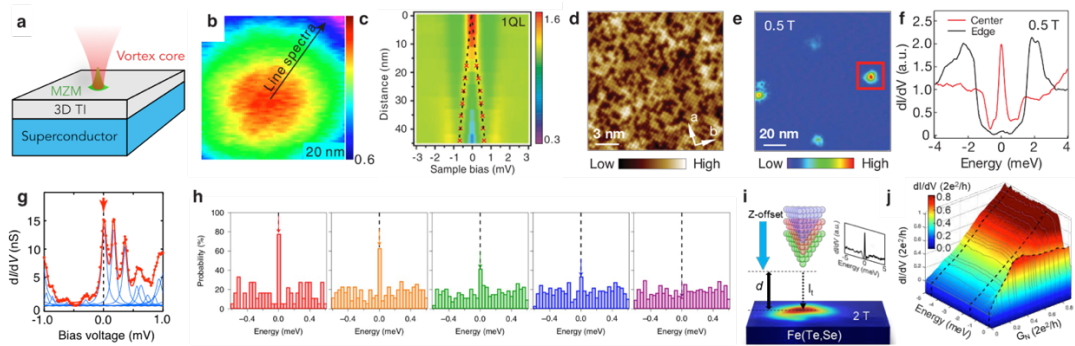

**Figure 3 | MZM vortex core states in 2D topological surface states a,** Schematics of a 3D TI bulk superconductor hybrid structure, in which a MZM is expected to localize as a ZBP in vortex cores on the surface. **b,** Zero bias conductance map recorded with an STM reveals a ZBP inside the vortex core of the $Bi_2Se_3$/$NbSe_2$ hybrid structure. **c,** Spectroscopic line cut along the black line in panel b, illustrating the spatial evolution of the ZBP. Panels b, c reproduced from[79]. **d,** STM topography of the $FeTe_{0.55}Se_{0.45}$ surface showing its atomically resolved square-lattice structure as well as the presence of surface impurities. **e,** A zero bias differential conductance map in the same area as d reveals vortices in the presence of an applied magnetic field of B=0.5 T. **f,** dI/dV point spectrum at the center of a vortex core (red rectangle in panel e) demonstrates the existence of a sharp ZBP.. **g,** Finite energy CdGM-states and a distinct ZBP appearing in the dI/dV-spectra recorded at the center of vortex cores on the $FeTe_{0.55}Se_{0.45}$ surface at a temperature of 80 mK. The red curve displays the experimental data and the blue curves correspond to fits for the peak analysis. **h** Probability for the occurrence of low energy states as a function of energy and for different applied magnetic fields ($B=\{1, 2, 3, 4, 6\}$ T, left to right). **i,** Schematics of an STM experiment to probe the conductance of electronic states occurring inside vortex cores on the surface of $FeTe_{0.55}Se_{0.45}$, e.g. the ZBP shown in the inset, as a function of tunnel junction transparency, by adjusting the tip-sample separation $d$. **j,** 3D-plot of the measured dI/dV-amplitude as a function of energy and tunnel contact transparency, $G_N$. Panels b, c reproduced from[79], panel d-f reproduced from[31], panels g,h reproduced from[80], panels i,j reproduced from[81].

An early proposal to realize TRS topological superconductivity is based on the 2D topological surface state of 3D TI[82,83], where proximity induced pairing of the helical surface Dirac electrons stabilizes a p+ip 2D topological superconducting phase[7]. A vortex core resulting from the application of an external magnetic field locally breaks TRS and represents a topological defect[84,85], which can host a MZM at its center (see Fig. 3a). First experimental efforts to realize this proposal focused on heterostructures of epitaxially grown $Bi_2Se_3$[83,86,87] thin films on the surface of superconducting bulk $NbSe_2$[88–90]. Spectroscopic measurements with the STM and complementing spin-resolved STM experiments on this material platform reported a zero bias peak inside the vortex cores at moderate magnetic fields (Fig. 3b), whose properties were consistent with the presence of a localized MZM[79,91–93]. However, spectroscopic imaging experiments revealed a continuous ZBP splitting into a pair of finite energy states over a range of 40 nm away from the vortex core (Fig. 3c). While such characteristics could result from other trivial low lying in-gap states[79,91,94], exhibiting a spatial distribution different from that of the ZBP, they contest the interpretation of the ZBP as a charge signature of a MZM. The interpretation of these sub-gap states is, additionally, complicated by a soft superconducting gap, resulting from a weak superconducting proximity effect in the quintuple layered structure of $Bi_2Se_3$[79,88].

In this regard, a major leap forward was achieved by the possibility of intrinsic superconductivity with a hard superconducting gap in the 2D topological surface states of $FeTe_{0.55}Se_{0.45}$ and $Li_{0.84}Fe_{0.16}OHFeSe$ [95,96]. Ensuing low-temperature STM experiments on the surface of these materials (Fig. 3d for $FeTe_{0.55}Se_{0.45}$) reported the observation of a sharp and spatially homogeneous ZBP inside the vortex core states (Fig. 3e, f)[31,96]. While the ZBP's properties, such as its spatial extent, spectral width, and temperature dependence have been reported to be consistent with those of a MZM, trivial vortex core states inside the superconducting gap, so-called Caroli-de Gennes-Matricon (CdGM) states[97,98], would exhibit similar characteristics; also see Sec. 4 'MZM spin signature'. Only a fraction of vortices were actually found to host a ZBP, while the others host CdGM states at finite energy. Hence, open questions about the origin of the ZBP and the role of the significant surface disorder remain. In the case of $FeTe_{0.55}Se_{0.45}$, disorder was also found to induce trivial surface states in some regions of the sample surface[99], and other experiments even reported the entire absence of any vortex core ZBP[100].

The debate on the origin of the ZBP could partially be resolved by a high-resolution study of the vortex core sub-gap states on the surface of $FeTe_{0.55}Se_{0.45}$, which was performed at a temperature of 80 mK[80]. This experiment, first, reproduced the experimental observation of vortex core ZBP. In addition, the trivial CdGM states appeared as distinct peaks at finite energy, well separated from the vortex core ZBP, see Fig. 3g. Interestingly, it was also reported that the relative occurrence of ZBP inside vortex cores decreases when the strength of an externally applied magnetic field increases. This intriguing observation, shown in Fig. 3h, hints at hybridization of MZM residing in neighboring vortices, by which their energy is shifted to finite energy[101].

Other studies addressed the origin of vortex core ZBP on the $Li_{0.84}Fe_{0.16}OHFeSe$ and $FeTe_{0.55}Se_{0.45}$ surfaces[81,102], by investigating their spectral properties as function of tip-sample distance, see Fig. 3i. Theory predicts that tunneling into a localized MZM via resonant Andreev reflection results in quantized conductance of the ZBP, which assumes universal value of the conductance quantum, $G_0 = \frac{2e^2}{h}$ ($e$ – elementary charge, $h$ – Planck's constant)[103]. This quantized conduction value would be independent of the tunnel junction conductance and yield a so-called conductance plateau, when the tip-sample distance is changed.

In the context of charge tunneling into a superconductor, it is instructive to generally consider that varying tip-sample distance changes the tunnel contact conductance[52] and can realize various tunnel regimes, where charge transport is dominated by different mechanisms. At comparably large tip-sample distances, such as those commonly used for topographic imaging and STS measurements, tunneling of individual quasiparticles dominates and gives rise to a superconducting gap, $\Delta$, and coherence peaks at energies, $|E|=\Delta$, in the tunneling spectrum, e.g. see Fig. 3f. At the same time, Cooper pair-breaking effects, which can result from finite temperature, magnetic adatoms or vortex cores for example, result in a finite quasiparticle density of states at $E<\Delta$. Such 'soft' gaps provide relaxation pathways for quasiparticles[104] and they facilitate tunneling into sub-gap states, such as MZM, Shiba states, and CdGM states, even at $E<\Delta$. In the presence of a 'hard' gap, such tunneling processes of individual quasiparticles would, by contrast, be suppressed. Considering the case of comparably small tip-sample distances, charge tunneling at $E<\Delta$ is dominated by Andreev reflections that give rise to rich spectral features inside the superconducting gap[105]. By injecting and extracting an electron and hole, respectively to form a Cooper pair, this tunnel process between tip and sample is mediated by Andreev bound states in the tunnel barrier and influences the spectral characteristics of charge tunneling into sub-gap states such as MZM, Shiba states, and CdGM-states[103,106].

Concerning tunneling into the vortex core ZBP on the $FeTe_{0.55}Se_{0.45}$ surface, the authors report that the ZBP amplitude is independent of tunnel contact conductance, when the tip-sample distance is decreased [81], see Fig. 3j. Similar observations of plateau-like behavior were reported for tunneling into one vortex core on the surface of $Li_{0.84}Fe_{0.16}OHFeSe$[102]. While the appearance of a conductance plateau at zero bias voltage is qualitatively consistent with resonant Andreev reflection into MZM[103], the reported plateau amplitude assumes non-universal values below the theoretically expected quantized value $G_0$ [81] or even exceeds $G_0$, when the tip is continuously lowered toward the sample surface[102]. These tunneling characteristics question their interpretation as signature of Andreev tunneling into a MZM. In fact, comparable experiments on quasiparticle tunneling into trivial Shiba states at $E<\Delta$ report similar results[104]. In that case, plateau behavior can be unambiguously assigned to a tunneling blockade due to a suppressed quasiparticle relaxation rate at low temperatures. Hence, the conductance plateau can be considered as a rather generic property of quasiparticle tunneling into localized sub-gap states, challenging its suitability as a tool to distinguish trivial from non-trivial zero energy states in STM experiments.

Beyond the search for MZM in the vortex cores of an externally applied magnetic field, the surface of $FeTe_{0.55}Se_{0.45}$ has also inspired other concepts in which a MZM could localize at structural defects. STM experiments on 1D line defects in monolayer $FeSe_{0.5}Te_{0.5}$ reported the observation of ZBP pairs, which are localized at the ends of 1D line defects[107], and the authors conjecture that these states could be the signatures of a Majorana Kramer's pair. Further STM experiments on structural domain walls, appearing on the surface of $FeSe_{0.45}Te_{0.55}$, observed a filling of the superconducting gap with an energetically flat quasiparticle density of states along the 1D domain wall trajectory[68]. Since such flat dI/dV amplitude inside the domain wall superconducting gap can arise from a linearly dispersing 1D quasiparticle state, this phenomenon is interpreted as a signature of a 1D chiral Majorana mode.

### 3.3 Majorana zero modes in one-dimensional topological edge states

Two-dimensional topological insulators (2D TI) host one-dimensional (1D) helical edge states[75,108,109]. It has been proposed that proximity-induced s-wave superconductivity in these states can realize a time-reversal symmetric topological superconducting phase. Its combination with segments of ferromagnetic material can realize an interface at which a localized MZM emerges[13,110,111]. Magnetism and superconductivity each induce their perspective gaps on the topological edge states, giving rise to a topological mass domain wall where the MZM is formed (see Fig. 4a). Several transport experiments have already reported superconducting pairing in the topological edge states of HgTe quantum well, InSb and bismuth (Bi) nano-wires[112–114]. Realizing a MZM in these materials using a domain wall with magnetism, however, has remained an open question to date, potentially due to the challenge that arises when coupling these states to suitable magnetic materials.

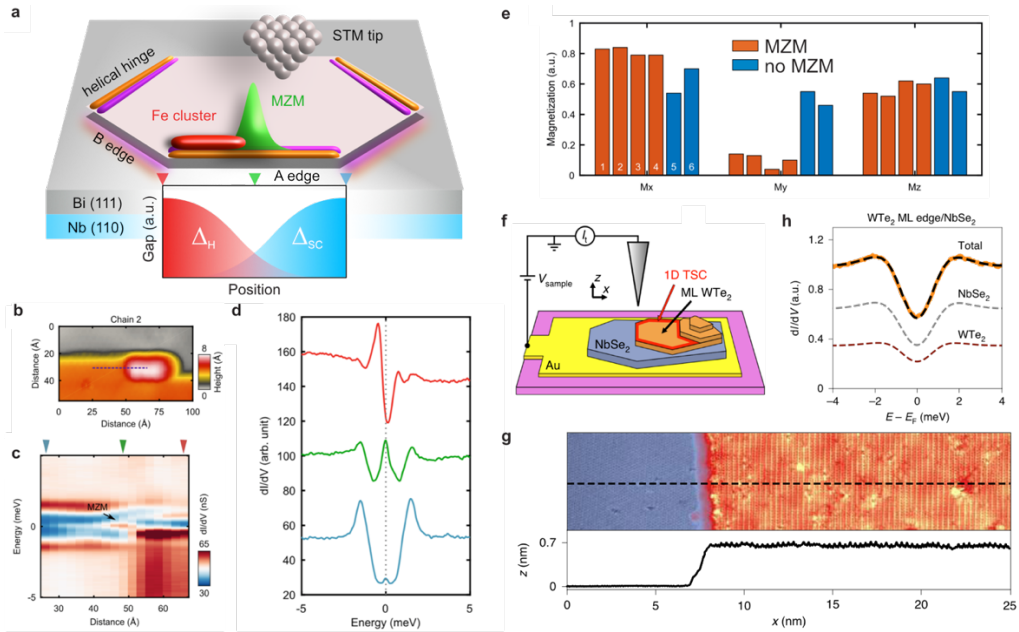

**Figure 4 | Topological superconductivity and MZM in 1D topological edge states. a** Schematic of a topological hybrid structure to realize MZM based on the helical hinge states of a Bi bilayer. Proximity-induced superconductivity $\Delta_{SC}$ is realized by realizing a Bi(111) thin film on top of the bulk superconductor Nb(110). A ferromagnetic Fe cluster adsorbed to the bilayer edge can induce a trivial magnetization gap $\Delta_H$. The topological mass domain realized at the interface between these regions localizes a MZM, which can be probed via local spectroscopy with an STM tip. **b** Close up STM topography of a Fe cluster decorating the bilayer edge. **c & d** A spectroscopic line cut along the black dashed line in **b** and individual point spectra demonstrate the emergence of a localized ZBP at the interface of the edge state with the Fe cluster. Panels a-d reproduced from[28]. **e** Magnetization components of several Fe clusters as reconstructed from spin-polarized STM experiments ($M_x$ points along the bilayer edge, $M_y$ perpendicular in-plane to the edge and $M_z$ out of plane). **f** Schematic of an STM experiment on a vdW-heterostructure assembled from NbSe$_2$ and 1T'-WTe$_2$ flakes deposited on a gold electrode, which serves as drain for the tunnel current. **g,** STM topography (top) and topographic line cut (bottom) of the edge of a WTe$_2$ monolayer flake on the surface of NbSe$_2$. **h,** The dI/dV-spectrum (yellow line) recorded on the WTe$_2$ monolayer edge displays a soft superconducting gap, which can be separated into dI/dV-contributions arising from the LDOS of WTe$_2$ edge state and the NbSe$_2$ substrate. Panels a-e reproduced from[28] and panels f-h reproduced from[115].

The topological edge state of bismuth (Bi)[78,116] is a promising alternative, because it appears on the edges of bilayers on the surface of Bi(111), where their properties can be explored with STM experiments[117–119]. Moreover, high quality Bi(111) thin films can be grown epitaxially[120], which has facilitated the observation of superconducting pairing inside the edge states of Bi(111) thin films grown on Nb(110)[28]. The challenge of inducing the topological mass domain wall for localizing a MZM was overcome by decorating the bilayer edges with self-assembled ferromagnetic Fe clusters (see Fig. 4a), which were found to induce a trivial magnetization gap inside the edge state band structure[28,119].

High-resolution spectroscopic measurements with the STM demonstrated the emergence of a localized ZBP at the Fe cluster-topological edge state interface (Fig. 4b-d), confirming the early theory proposals for the emergence of MZM on such a platform[13,110,111]. Additionally, the presence of the ZBP showed a characteristic dependence on the Fe cluster magnetization, see Fig. 4e, justifying its interpretation as a charge signature of a MZM[28]. We note that this observation could provide avenues to manipulate MZM with nanoscale magnetic switches, where the cluster magnetization could be tuned with external magnetic fields or spin-polarized currents form the STM tip. While the reported ZBP properties were consistent with results from model calculations, only one cluster-edge state interface at the so-called A edge (cf. Fig. 4a) was reported to show a prominent ZBP, whereas the other interface of the cluster with the B edge revealed an enhanced zero bias conductance. This observation can be accounted for by the hybridization of the topological edge state with the bulk states along the B edge. However, further experiments are desirable, in which the ferromagnetic cluster decorates the center of an A edge, and a pair of distinct ZBPs emerges on both sides of the cluster.

Also in the context of 2D topological insulators, vdW-heterostructures have been established as an attractive new platform to study proximity-induced s-wave superconductivity with large topological gaps in the 1D helical edge state of monolayer transition-metal-dichalcogenides[77]. Superconducting pairing in the topological edge state of monolayer 1T'-WTe2[121,122] has been realized, by depositing a micrometer-sized flake of this compound on top of the surface of bulk NbSe2, see Fig. 4f.[115] Spectroscopic measurements with the STM on the edge of this flake revealed the presence of a superconducting gap in the LDOS of the topological edge state, see Fig. 4g,h. This observation opens an attractive avenue to explore topological superconductivity and MZM in experiments on other heterostructures, which include flakes of 2D ferromagnets[67,123,124] or local gates with the potential to engender topological superconductivity through the intrinsic proximity effect[125].

### 4. How to distinguish trivial from topological ZBP

Our review illustrates that ZBP have been sighted in STM experiments across a variety of material platforms, which have been proposed to realize various theoretical concepts for topological superconductivity. While several control experiments, such the suppression of the ZBP in the absence of superconductivity[24,58] or studies of the ZBP's temperature and magnetic field dependence[31,79,96,107], are commonly performed, the interpretation of the observed ZBP as a charge signature of a MZM rests on the assumption that a ZBP is detected, when the parameters of the system make it most likely to be in a topological superconducting phase. An accidental trivial zero-energy state, such as a Shiba state fine-tuned to zero bias, will, however, result in experimental characteristics similar to those of a MZM. Sighting of a ZBP on a suitable material platform by itself does not, therefore, constitute sufficient evidence to justify its interpretation as a charge signature of a MZM. STM experiments with functional tips have the ability probe other predicted MZM properties, such as its spin and electron-hole symmetry, and they were established as a way out of this predicament to distinguish trivial from topological ZBP, as we will discuss in the following.

**4.1 MZM electron-hole symmetry**

The intrinsic electron-hole symmetry of a MZM, which is imposed by its Bogoliubov-quasiparticle character[126], can be probed in tunnel spectroscopy experiments with superconducting STM tips[47]. The presence of a superconducting tip gap, $\Delta_T$, facilitates a separate measurement of the electron and hole sectors of a ZBP at different energies, by shifting their spectral weight to finite voltages. An electron-hole symmetric ZBP will be mapped to a pair of peaks with equal amplitude appearing at $V=\pm\Delta_T/e$. While thermally excited resonances can render the detection of this symmetry challenging[25], high-resolution STS performed at dilution-refrigerator temperatures observed such electron-hole symmetry of the ZBP at the end of the ferromagnetic Fe chain on Pb(110)[45]. In these experiments, trivial Shiba states appeared as pairs of asymmetric peaks at $V>\Delta_T$, reflecting their intrinsic electron-hole asymmetry[54], and a pair of peaks with equal amplitude was measured at $\Delta_T$. This observation further consolidated the MZM interpretation of the ZBP on the

Fe chain platform[24–26,45], and underscored the potential of functional STM tips for the study of MZM.

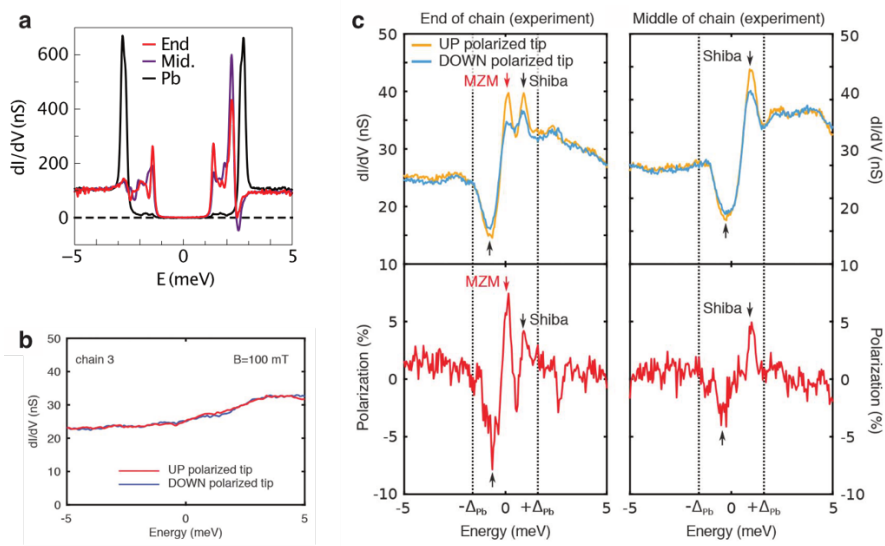

**Figure 5 | Detection of MZM using superconducting and spin polarized STM spectroscopy. a,** Individual dI/dV spectra, which were recorded with a superconducting STM tip on the bare Pb (black line) substrate, at the Fe chain center (Mid., purple line), and at the Fe chain end (red line). The end spectrum shows electron-hole symmetric peak amplitude at $e|V| = \Delta_t$. **b,** Spectra measured on a Fe chain at 100 mT with UP (red) and DOWN (blue) polarized tips, showing the compensation of the unequal spin densities of the ferromagnetic Fe chain, cf. Fig.1e, in the dI/dV-spectrum by the STM set-point effect. **d,** Experimentally obtained spectra at the end of chain (left column) and at the center of chain (right column) and their corresponding polarization. Yellow and blue curves are taken with UP and DOWN polarized tips, respectively. Red arrows mark the zero-energy end state and black arrows mark the van-hove singularity of the Shiba band. Panel a reproduced from[45] and panels b,c reproduced from[46].

## 4.2 MZM spin signature

Theoretical investigations have proposed that the MZM can leave unique fingerprints in spin-sensitive measurements, therefore providing the means to distinguish topological MZM from a trivial zero energy state. [92,127–132]. STM experiments with magnetic STM tips, also called spin-polarized spectroscopy (SPS), can measure a sample's spin-polarization with atomic resolution[133], and they are particularly well-suited for these kind of studies. When tunneling occurs between a ferromagnetic tip and sample, respectively, the tunnel conductance at given bias voltage will depend on their relative magnetization direction. This effect is also called tunnel-magneto resistance, and it allows to calculate the spin-polarization of the tunnel current, which gives direct access to a sample's magnetic properties[133]. In case the distance between tip and sample is adjusted so as to maintain a constant current set-point, the spin-dependent contribution to the tunnel current, which arises from unequal normal state spin densities, $\rho_N^{\uparrow/\downarrow}$, in a sample, is compensated (see Fig. 5b); this is also referred to as set-point effect. The same measurement in the superconducting state opens a possibility to detect any spin polarization of the in-gap states beyond that caused by the ferromagnetism of the atomic chains in the normal state.

Model calculations of spin-dependent tunneling into localized sub-gap states of a magnetic impurity on a superconducting surface reveal that the spin-polarization of a pair of Shiba states at $\pm E$ is asymmetric about zero energy and that the Shiba state spin-density of states is bound by $\rho_N^{\uparrow/\downarrow}$ outside the superconducting gap[54]. Accordingly, a trivial Shiba state tuned to zero energy is expected to show no spin-polarization in SPS experiments, because the spin densities are offset by the set-point effect of the tunnel current[44]. By contrast, the MZM spin-densities are, in first order, only dependent on the magnetic exchange interaction, and a MZM is expected to show a finite spin-polarization despite the set-point effect. Hence, a topological ZBP can be distinguished from a trivial ZBP by measuring its spin-polarization in SPS experiments[44].

Jeon *et al*. confirmed this hypothesis by performing spin-polarized STM experiments on the sub-gap states appearing in Fe chains on Pb using Fe coated Cr tips[46]. Consistent with the model calculations, the Shiba states within the chain exhibit an energy-asymmetric spin polarization. Comparable results were also reported in similar experiments on Co chains on Pb(110)[60] and single magnetic impurities[134]. More importantly, additional measurements at the Fe chain end revealed a distinct spin-polarized signal of the ZBP. By virtue of the above considerations, this observation directly demonstrates its non-trivial origin and firmly establishes the MZM interpretation of the ZBP in the Fe chain on Pb(110) platform. Similar SPS measurements have also been employed to confirm the MZM nature of the ZBP observed in the proximitized topological edge state of Bi[28]. The observed strong spin-polarization of the ZBP was found to be of opposite sign to that of the Shiba state on top of the Fe cluster. This finding is consistent with analytical and numerical model calculations, which predict that MZM and Shiba state spins point along different spatial axes.

Additional results from recent spin-polarized STM experiments on Fe adatoms adsorbed to the surface of $FeSe_{0.45}Te_{0.55}$ underscore the relevance of diagnostic tools that probe other MZM properties than just the presence of a ZBP by itself. Initial spectroscopic measurements on top of the adatoms revealed the presence of a sharp ZBP[33], which were interpreted as signatures of MZM occurring inside a quantum anomalous vortex core[135]. We note that similar results and conclusions were also reported for Fe adatoms deposited on the surface LiFeAs and PbTaSe2[136]. However, a recent study using spin-polarized measurements by Wang et al. experimentally demonstrated the absence of finite spin-polarization for the ZBP observed on top of Fe adatoms on $FeSe_{0.45}Te_{0.55}$[34]. The conflicting outcome of these two studies clearly precludes the interpretation of the ZBP on this platform as a signature of a MZM. More broadly, this result emphasizes that tracking the temperature and field-dependence of a ZBP alone does not provide sufficient information to draw conclusions on its topological origin.

Other auspicious proposals to probe the MZM origin of ZBP with STM experiments exist. They involve the use of microwave radiation coupled to the tunnel junction for accurately testing spectral properties of ZBP[137,138], and a shot noise analysis of the tunnel current to discriminate tunneling into a MZM from tunneling into trivial states[139]. Finally, we anticipate that Josephson STM measurements[140], which use a superconducting STM tip, could present a valuable tool to scrutinize the presence of topological p-wave superconducting pairing in the bulk of 1D and 2D systems[29,32,58,61]. Josephson STM focuses on the measurement of a Cooper pair current near zero bias voltage, through which it can quantify the amplitude of the superconducting order parameter with atomic resolution[141,142]. Since tunneling of Cooper pairs between spin-singlet (s-wave) superconductivity in the STM tip and spin-triplet (p-wave) superconductivity in the sample cannot occur, Cooper pair current would be locally suppressed. Such local variations can be detected with Josephson STM and would yield a clear experimental signature for p-wave superconductivity in 1D and 2D systems.

## 5. Outlook and potential future experiments

The diversity of results from STM experiments presented in this review article illustrate the significant role STM has played for exploring topological superconductivity and visualizing the presence of MZM across a variety of material platforms[24,28–32,58,61,68,79,115,143]. More recently, a variety of STM experiments also succeeded in establishing novel material platforms that lend themselves to the integration into electron transport experiments; noteworthy recent examples of such material platforms include the hinge state of bismuth, whose properties can be studied in quantum transport experiments based on bismuth nano-ribbons[28,114], and the topological edge state of monolayer 1T-WTe$_2$ in vdW-heterostructures devices[115,125,144]. These developments close the gap between microscopic studies of MZM properties with the STM and measurements of global material properties in device transport, and they provide avenues to explore more complex MZM braiding experiments in the future[13,111,145].

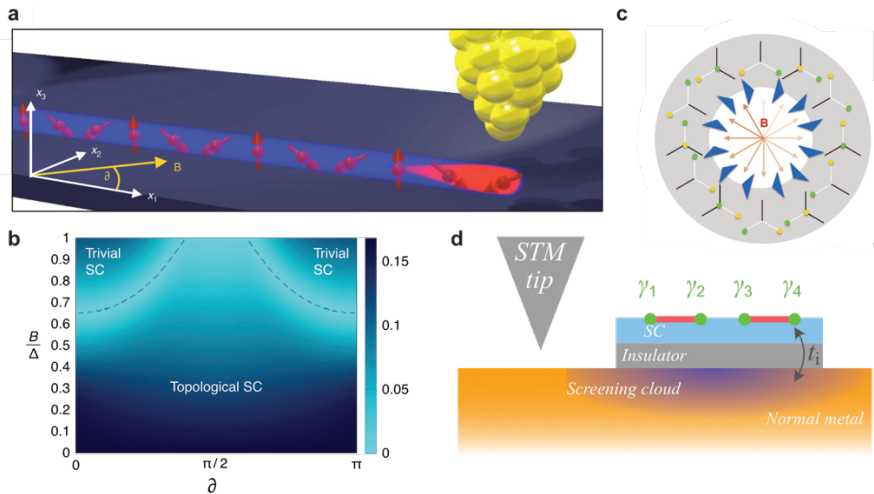

**Figure 6 | MZM manipulation detected with the STM a,** Schematics of a STM experiments on a Fe chain with helical spin-order to probe the emergence of a MZM end state as a function of the in-plane angle, $\partial$, between the chain and magnetic field axes, respectively. **b,** Topological phase diagram of the experimental setup shown in panel a. The amplitude, B, with respect to the superconducting gap, $\Delta$, and the in-plane angle, $\partial$, of the magnetic field can tune between topological and trivial electronic phases. **c** Schematics of an experiment to study MZM braiding on Fe helical chains by applying a rotating in-plane magnetic field to Y-junction geometry of a magnetic chain with helical spin order. Panels a-c reproduced from[48]. **d** Coupling two pairs of MZM on a superconducting island deposited on a metallic substrate can induce a topological Kondo effect, the screening cloud of which could be detected in STS experiments.

At the same time, theoretical studies of the atomic chain platform proposed STM measurements to test the two essential properties of MZM that are key to their application as topologically protected qubits—their non-Abelian exchange statistics and anyonic character. This outlines the future direction of research on MZM with STM experiments.

The possibly simplest demonstration of controlling MZM occurring at the ends of magnetic atomic chains on a superconducting surface is the study of their properties as function of the relative angle of an in-plane magnetic field to that of the chain direction, see Fig.6a. Theory[48] predicts that a spin chain with helical magnetic order would be driven out of the topological phase as a function of this angle, see Fig.6b. Hence, in a situation where the chain displays a semicircle-like geometry, trivial and topological segments can coexist, and the MZM would be localized at the boundary between these phases along the chain, depending on the local angle of the field with respect to the chain. Spatially resolved STM spectroscopy as function of position along the chain and the applied magnetic field at different angles, would enable us to examine this approach for manipulating the chain topology.

An extension of this proposal considers the properties of a T-junction chain as a function of a rotating in-plane magnetic field[48]. Examining the influence of the magnetic field on a tri-junction geometry with a 120º angle, theoretical analyses predict that the magnetic field rotation can drive each arm of such tri-junctions sequentially in and out of the topological phase. As shown in Fig. 6c, at each orientation of the applied field, precisely one pair of MZM appears at the ends of two out of three chains in the tri-junction and can be detected by their ZBP at respective locations. The process by which MZMs in these tri-junctions are fused and created sequentially as a function of field rotation results in their exchange at a rotation angle of $\pi$ and a single braid at $2\pi$. A real space manipulation of a MZM pair, following this pattern, corresponds to a braiding process, which is at the heart of the complex physics of MZMs and could be visualized in STS measurements with the STM.

We here propose that chains of magnetic atoms may also serve as a suitable testbed to demonstrate the ground state degeneracy of a system containing more than two MZM. Theoretical analyses of MZM quantum dot devices propose that coupling three (out of two pairs of) degenerate MZM to metallic leads constitutes an effective spin-1 particle, which gives rise to the topological Kondo effect[146]. In a similar fashion, we envision an atomic scale platform with two atomic chains placed on a superconducting film, each of which hosts a pair of MZM, see Fig.6d. If a thin electrically insulating layer isolates the superconductor from a normal metallic substrate forming a Coulomb island, we conjecture that the coupling, $t_i$, of the MZM with the conduction electrons in the substrate underneath could mediate a Kondo-type interaction governed by $t_i$ and the Coulomb energy, displaying non-Fermi liquid behavior. The resulting Kondo screening cloud, which arises in the metal substrate surrounding the island, can be detected in STS measurements[147], see Fig.6d. The emergence of the topological Kondo effect requires the existence of more than two degenerate MZM. The experimental detection of Kondo cloud would, therefore, provide direct evidence for the anyonic character of MZM. We envision that such a sample platform could be realized on the basis of vdW-heterostructures, by assembling mono- and few-layer flakes of suitable materials, e.g. superconducting monolayer NbSe$_2$, insulating h-BN and metallic Graphene[66].

The recent demonstration of atom by atom assembly of magnetic nanostructures on the surface of different superconducting substrates with an STM tip[30,49,50] makes the possibility of these studies more likely. These experiments are rather challenging as they require obtaining atomically clean superconducting surfaces[30,148] that permit atomic assembly of elaborate atomic structures, and the right material combination to realize 1D topological superconductivity with MZM end states. These developments will, ultimately, be driven by the realization of novel material platforms and concepts, such as those based on higher-order topological superconductors[149–151] or quantum-spin liquids[152–154]. Nevertheless, STM experiments, combining the unique ability to visualize, investigate, and potentially manipulate MZM with atomic scale precision, will continue to be at the forefront of research on topological superconductivity and MZM in condensed matter systems.

# References

1. Kitaev, A. Y. Unpaired Majorana fermions in quantum wires. *Physics-Uspekhi* **44**, 131 (2001).
2. Kitaev, A. Y. Fault-tolerant quantum computation by anyons. *Ann. Phys. (N. Y)*. **303**, 2–30 (2003).
3. Nayak, C., Simon, S. H., Stern, A., Freedman, M. & Sarma, S. Das. Non-Abelian anyons and topological quantum computation. *Rev. Mod. Phys.* **80**, 1083 (2008).
4. Stern, A. & Lindner, N. H. Topological quantum computation—from basic concepts to first experiments. *Science (80-. )*. **339**, 1179–1184 (2013).
5. Lahtinen, V. & Pachos, J. K. A short introduction to topological quantum computation. *SciPost Phys.* **3**, (2017).
6. Li, J. *et al.* Topological superconductivity induced by ferromagnetic metal chains. *Phys. Rev. B* **90**, 235433 (2014).
7. Fu, L. & Kane, C. L. Superconducting proximity effect and Majorana fermions at the surface of a topological insulator. *Phys. Rev. Lett.* **100**, 96407 (2008).
8. Nakosai, S., Tanaka, Y. & Nagaosa, N. Two-dimensional p-wave superconducting states with magnetic moments on a conventional s-wave superconductor. *Phys. Rev. B* **88**, 180503 (2013).
9. Nadj-Perge, S., Drozdov, I. K., Bernevig, B. A. & Yazdani, A. Proposal for realizing Majorana fermions in chains of magnetic atoms on a superconductor. *Phys. Rev. B* **88**, 20407 (2013).
10. Klinovaja, J., Stano, P., Yazdani, A. & Loss, D. Topological superconductivity and Majorana fermions in RKKY systems. *Phys. Rev. Lett.* **111**, 186805 (2013).
11. Pientka, F., Glazman, L. I. & von Oppen, F. Topological superconducting phase in helical Shiba chains. *Phys. Rev. B* **88**, 155420 (2013).
12. Pientka, F., Glazman, L. I. & von Oppen, F. Unconventional topological phase transitions in helical Shiba chains. *Phys. Rev. B* **89**, 180505 (2014).
13. Fu, L. & Kane, C. L. Josephson current and noise at a superconductor/quantum-spin-Hall-insulator/superconductor junction. *Phys. Rev. B* **79**, 161408 (2009).
14. Lutchyn, R. M., Sau, J. D. & Sarma, S. Das. Majorana fermions and a topological phase transition in semiconductor-superconductor heterostructures. *Phys. Rev. Lett.* **105**, 77001 (2010).
15. Oreg, Y., Refael, G. & Von Oppen, F. Helical liquids and Majorana bound states in quantum wires. *Phys. Rev. Lett.* **105**, 177002 (2010).
16. Duckheim, M. & Brouwer, P. W. Andreev reflection from noncentrosymmetric superconductors and Majorana bound-state generation in half-metallic ferromagnets. *Phys. Rev. B* **83**, 54513 (2011).
17. Chung, S. B., Zhang, H.-J., Qi, X.-L. & Zhang, S.-C. Topological superconducting phase and Majorana fermions in half-metal/superconductor heterostructures. *Phys. Rev. B* **84**, 60510 (2011).
18. Potter, A. C. & Lee, P. A. Topological superconductivity and Majorana fermions in metallic surface states. *Phys. Rev. B* **85**, 94516 (2012).
19. Braunecker, B. & Simon, P. Interplay between classical magnetic moments and superconductivity in quantum one-dimensional conductors: toward a self-sustained topological Majorana phase. *Phys. Rev. Lett.* **111**, 147202 (2013).

20. Vazifeh, M. M. & Franz, M. Self-organized topological state with Majorana fermions. *Phys. Rev. Lett.* **111**, 206802 (2013).
21. Mourik, V. *et al.* Signatures of Majorana fermions in hybrid superconductor-semiconductor nanowire devices. *Science (80-. ).* **336**, 1003–1007 (2012).
22. Das, A. *et al.* Zero-bias peaks and splitting in an Al--InAs nanowire topological superconductor as a signature of Majorana fermions. *Nat. Phys.* **8**, 887–895 (2012).
23. Deng, M. T. *et al.* Majorana bound state in a coupled quantum-dot hybrid-nanowire system. *Science (80-. ).* **354**, 1557–1562 (2016).
24. Nadj-Perge, S. *et al.* Observation of Majorana fermions in ferromagnetic atomic chains on a superconductor. *Science (80-. ).* **346**, 602–607 (2014).
25. Ruby, M. *et al.* End States and Subgap Structure in Proximity-Coupled Chains of Magnetic Adatoms. *Phys. Rev. Lett.* **115**, 197204 (2015).
26. Pawlak, R. *et al.* Probing atomic structure and Majorana wavefunctions in mono-atomic Fe chains on superconducting Pb surface. *npj Quantum Inf.* **2**, 16035 (2016).
27. Yazdani, A., da Silva Neto, E. H. & Aynajian, P. Spectroscopic Imaging of Strongly Correlated Electronic States. *Annu. Rev. Condens. Matter Phys.* **7**, 11–33 (2016).
28. Jäck, B. *et al.* Observation of a Majorana zero mode in a topologically protected edge channel. *Science (80-. ).* **364**, 1255 LP – 1259 (2019).
29. Ménard, G. C. *et al.* Two-dimensional topological superconductivity in Pb/Co/Si(111). *Nat. Commun.* **8**, 2040 (2017).
30. Kim, H. *et al.* Toward tailoring Majorana bound states in artificially constructed magnetic atom chains on elemental superconductors. *Sci. Adv.* **4**, eaar5251 (2018).
31. Wang, D. *et al.* Evidence for Majorana bound states in an iron-based superconductor. *Science (80-. ).* **362**, 333 LP – 335 (2018).
32. Manna, S. *et al.* Signature of a pair of Majorana zero modes in superconducting gold surface states. *Proc. Natl. Acad. Sci.* **117**, 8775 LP – 8782 (2020).
33. Yin, J.-X. *et al.* Observation of a robust zero-energy bound state in iron-based superconductor Fe(Te,Se). *Nat. Phys.* **11**, 543–546 (2015).
34. Wang, D., Wiebe, J., Zhong, R., Gu, G. & Wiesendanger, R. Spin-polarized Yu-Shiba-Rusinov states in an iron based superconductor. in (2020).
35. Pan, H. & Das Sarma, S. Physical mechanisms for zero-bias conductance peaks in Majorana nanowires. *Phys. Rev. Res.* **2**, 13377 (2020).
36. Pan, H., Cole, W. S., Sau, J. D. & Das Sarma, S. Generic quantized zero-bias conductance peaks in superconductor-semiconductor hybrid structures. *Phys. Rev. B* **101**, 24506 (2020).
37. Vaitiekenas, S. *et al.* Flux-induced topological superconductivity in full-shell nanowires. *Science (80-. ).* **367**, (2020).
38. Valentini, M. *et al.* Flux-tunable Andreev bound states in hybrid full-shell nanowires. *arXiv Prepr. arXiv2008.02348* (2020).
39. Frolov, S. M., Manfra, M. J. & Sau, J. D. Topological superconductivity in hybrid devices. *Nat. Phys.* **16**, 718–724 (2020).
40. Yu, P. *et al.* Non-Majorana states yield nearly quantized conductance in proximatized nanowires. *Nat. Phys.* (2021). doi:10.1038/s41567-020-01107-w
41. Zhang, H. *et al.* Quantized Majorana conductance. *Nature* **556**, 74–79 (2018).
42. Zhang, H. *et al.* Editorial Expression of Concern: Quantized Majorana conductance. *Nature* **581**, E4–E4 (2020).

43. Zhang, H. *et al.* Retraction Note: Quantized Majorana conductance. *Nature* (2021). doi:10.1038/s41586-021-03373-x
44. Li, J., Jeon, S., Xie, Y., Yazdani, A. & Bernevig, B. A. Majorana spin in magnetic atomic chain systems. *Phys. Rev. B* **97**, 125119 (2018).
45. Feldman, B. E. *et al.* High-resolution studies of the Majorana atomic chain platform. *Nat. Phys.* **13**, 286–291 (2017).
46. Jeon, S. *et al.* Distinguishing a Majorana zero mode using spin-resolved measurements. *Science (80-. ).* **358**, 772 LP – 776 (2017).
47. Peng, Y., Pientka, F., Vinkler-Aviv, Y., Glazman, L. I. & von Oppen, F. Robust Majorana Conductance Peaks for a Superconducting Lead. *Phys. Rev. Lett.* **115**, 266804 (2015).
48. Li, J., Neupert, T., Bernevig, B. A. & Yazdani, A. Manipulating Majorana zero modes on atomic rings with an external magnetic field. *Nat. Commun.* **7**, 10395 (2016).
49. Odobesko, A. *et al.* Observation of tunable single-atom Yu-Shiba-Rusinov states. *Phys. Rev. B* **102**, 174504 (2020).
50. Schneider, L., Beck, P., Wiebe, J. & Wiesendanger, R. Atomic-scale spin-polarization maps using functionalized superconducting probes. *arXiv Supercond.* (2020).
51. Ding, H. *et al.* Tuning interactions between spins in a superconductor. *to appear* (2021).
52. Chen, C. J. *Introduction to Scanning Tunneling Microscopy: Second Edition. Monographs on the Physics and Chemistry of Materials* (Oxford University Press, 2007). doi:10.1093/acprof:oso/9780199211500.001.0001
53. Yazdani, A., Jones, B. A., Lutz, C. P., Crommie, M. F. & Eigler, D. M. Probing the Local Effects of Magnetic Impurities on Superconductivity. *Science (80-. ).* **275**, 1767 LP – 1770 (1997).
54. Balatsky, A. V, Vekhter, I. & Zhu, J.-X. Impurity-induced states in conventional and unconventional superconductors. *Rev. Mod. Phys.* **78**, 373–433 (2006).
55. Christensen, M. H., Schecter, M., Flensberg, K., Andersen, B. M. & Paaske, J. Spiral magnetic order and topological superconductivity in a chain of magnetic adatoms on a two-dimensional superconductor. *Phys. Rev. B* **94**, 144509 (2016).
56. Braunecker, B., Japaridze, G. I., Klinovaja, J. & Loss, D. Spin-selective Peierls transition in interacting one-dimensional conductors with spin-orbit interaction. *Phys. Rev. B* **82**, 45127 (2010).
57. Peng, Y., Pientka, F., Glazman, L. I. & von Oppen, F. Strong Localization of Majorana End States in Chains of Magnetic Adatoms. *Phys. Rev. Lett.* **114**, 106801 (2015).
58. Kezilebieke, S. *et al.* Topological superconductivity in a van der Waals heterostructure. *Nature* **588**, 424–428 (2020).
59. Rainis, D. & Loss, D. Majorana qubit decoherence by quasiparticle poisoning. *Phys. Rev. B* **85**, 174533 (2012).
60. Ruby, M., Heinrich, B. W., Peng, Y., von Oppen, F. & Franke, K. J. Exploring a Proximity-Coupled Co Chain on Pb(110) as a Possible Majorana Platform. *Nano Lett.* **17**, 4473–4477 (2017).
61. Palacio-Morales, A. *et al.* Atomic-scale interface engineering of Majorana edge modes in a 2D magnet-superconductor hybrid system. *Sci. Adv.* **5**, eaav6600 (2019).
62. Röntynen, J. & Ojanen, T. Topological Superconductivity and High Chern Numbers in 2D Ferromagnetic Shiba Lattices. *Phys. Rev. Lett.* **114**, 236803 (2015).
63. Li, J. *et al.* Two-dimensional chiral topological superconductivity in Shiba lattices. *Nat. Commun.* **7**, 12297 (2016).

64. LaShell, S., McDougall, B. A. & Jensen, E. Spin Splitting of an Au(111) Surface State Band Observed with Angle Resolved Photoelectron Spectroscopy. *Phys. Rev. Lett.* **77**, 3419–3422 (1996).
65. Dean, C. R. *et al.* Boron nitride substrates for high-quality graphene electronics. *Nat. Nanotechnol.* **5**, 722–726 (2010).
66. Geim, A. K. & Grigorieva, I. V. Van der Waals heterostructures. *Nature* **499**, 419–425 (2013).
67. Chen, W. *et al.* Direct observation of van der Waals stacking–dependent interlayer magnetism. *Science (80-. ).* **366**, 983 LP – 987 (2019).
68. Wang, Z. *et al.* Evidence for dispersing 1D Majorana channels in an iron-based superconductor. *Science (80-. ).* **367**, 104 LP – 108 (2020).
69. Hasan, M. Z. & Kane, C. L. Colloquium: Topological insulators. *Rev. Mod. Phys.* **82**, 3045–3067 (2010).
70. Beenakker, C. W. J. Search for Majorana Fermions in Superconductors. *Annu. Rev. Condens. Matter Phys.* **4**, 113–136 (2013).
71. Akhmerov, A. R., Nilsson, J. & Beenakker, C. W. J. Electrically Detected Interferometry of Majorana Fermions in a Topological Insulator. *Phys. Rev. Lett.* **102**, 216404 (2009).
72. Tanaka, Y., Yokoyama, T. & Nagaosa, N. Manipulation of the Majorana Fermion, Andreev Reflection, and Josephson Current on Topological Insulators. *Phys. Rev. Lett.* **103**, 107002 (2009).
73. Linder, J., Tanaka, Y., Yokoyama, T., Sudbø, A. & Nagaosa, N. Unconventional Superconductivity on a Topological Insulator. *Phys. Rev. Lett.* **104**, 67001 (2010).
74. Aguado, R. Majorana quasiparticles in condensed matter. *Riv. Del Nuovo Cim.* **40**, 523–593 (2017).
75. Bernevig, B. A., Hughes, T. L. & Zhang, S.-C. Quantum Spin Hall Effect and Topological Phase Transition in HgTe Quantum Wells. *Science (80-. ).* **314**, 1757 LP – 1761 (2006).
76. Liu, C., Hughes, T. L., Qi, X.-L., Wang, K. & Zhang, S.-C. Quantum Spin Hall Effect in Inverted Type-II Semiconductors. *Phys. Rev. Lett.* **100**, 236601 (2008).
77. Qian, X., Liu, J., Fu, L. & Li, J. Quantum spin Hall effect in two-dimensional transition metal dichalcogenides. *Science (80-. ).* **346**, 1344 LP – 1347 (2014).
78. Schindler, F. *et al.* Higher-order topology in bismuth. *Nat. Phys.* **14**, 918–924 (2018).
79. Xu, J.-P. *et al.* Experimental Detection of a Majorana Mode in the core of a Magnetic Vortex inside a Topological Insulator-Superconductor ${\mathrm{Bi}}_{2}{\mathrm{Te}}_{3}/{\mathrm{NbSe}}_{2}$ Heterostructure. *Phys. Rev. Lett.* **114**, 17001 (2015).
80. Machida, T. *et al.* Zero-energy vortex bound state in the superconducting topological surface state of Fe(Se,Te). *Nat. Mater.* **18**, 811–815 (2019).
81. Zhu, S. *et al.* Nearly quantized conductance plateau of vortex zero mode in an iron-based superconductor. *Science (80-. ).* **367**, 189 LP – 192 (2020).
82. Fu, L., Kane, C. L. & Mele, E. J. Topological Insulators in Three Dimensions. *Phys. Rev. Lett.* **98**, 106803 (2007).
83. Xia, Y. *et al.* Observation of a large-gap topological-insulator class with a single Dirac cone on the surface. *Nat. Phys.* **5**, 398–402 (2009).
84. Read, N. & Green, D. Paired states of fermions in two dimensions with breaking of parity and time-reversal symmetries and the fractional quantum Hall effect. *Phys. Rev. B* **61**, 10267–10297 (2000).

85. Ivanov, D. A. Non-Abelian Statistics of Half-Quantum Vortices in $\mathit{p}$-Wave Superconductors. *Phys. Rev. Lett.* **86**, 268–271 (2001).
86. Zhang, H. et al. Topological insulators in Bi2Se3, Bi2Te3 and Sb2Te3 with a single Dirac cone on the surface. *Nat. Phys.* **5**, 438–442 (2009).
87. Hsieh, D. et al. A tunable topological insulator in the spin helical Dirac transport regime. *Nature* **460**, 1101–1105 (2009).
88. Wang, M.-X. et al. The Coexistence of Superconductivity and Topological Order in the $Bi_2Se_3$ Thin Films. *Science (80-. ).* **336**, 52 LP – 55 (2012).
89. Xu, S.-Y. et al. Momentum-space imaging of Cooper pairing in a half-Dirac-gas topological superconductor. *Nat. Phys.* **10**, 943–950 (2014).
90. Xu, J.-P. et al. Artificial Topological Superconductor by the Proximity Effect. *Phys. Rev. Lett.* **112**, 217001 (2014).
91. Kawakami, T. & Hu, X. Evolution of Density of States and a Spin-Resolved Checkerboard-Type Pattern Associated with the Majorana Bound State. *Phys. Rev. Lett.* **115**, 177001 (2015).
92. Hu, L.-H., Li, C., Xu, D.-H., Zhou, Y. & Zhang, F.-C. Theory of spin-selective Andreev reflection in the vortex core of a topological superconductor. *Phys. Rev. B* **94**, 224501 (2016).
93. Sun, H.-H. et al. Majorana Zero Mode Detected with Spin Selective Andreev Reflection in the Vortex of a Topological Superconductor. *Phys. Rev. Lett.* **116**, 257003 (2016).
94. Chiu, C.-K., Gilbert, M. J. & Hughes, T. L. Vortex lines in topological insulator-superconductor heterostructures. *Phys. Rev. B* **84**, 144507 (2011).
95. Zhang, P. et al. Observation of topological superconductivity on the surface of an iron-based superconductor. *Science (80-. ).* **360**, 182 LP – 186 (2018).
96. Liu, Q. et al. Robust and Clean Majorana Zero Mode in the Vortex Core of High-Temperature Superconductor $\mathbf{(}{\mathrm{Li}}_{0.84}{\mathrm{Fe}}_{0.16}\mathbf{)}\mathrm{OHFeSe}$. *Phys. Rev. X* **8**, 41056 (2018).
97. Caroli, C., De Gennes, P. G. & Matricon, J. Bound Fermion states on a vortex line in a type II superconductor. *Phys. Lett.* **9**, 307–309 (1964).
98. Hess, H. F., Robinson, R. B., Dynes, R. C., Valles, J. M. & Waszczak, J. V. Scanning-Tunneling-Microscope Observation of the Abrikosov Flux Lattice and the Density of States near and inside a Fluxoid. *Phys. Rev. Lett.* **62**, 214–216 (1989).
99. Kong, L. et al. Half-integer level shift of vortex bound states in an iron-based superconductor. *Nat. Phys.* **15**, 1181–1187 (2019).
100. Chen, M. et al. Discrete energy levels of Caroli-de Gennes-Matricon states in quantum limit in FeTe0.55Se0.45. *Nat. Commun.* **9**, 970 (2018).
101. Chiu, C.-K., Machida, T., Huang, Y., Hanaguri, T. & Zhang, F.-C. Scalable Majorana vortex modes in iron-based superconductors. *Sci. Adv.* **6**, eaay0443 (2020).
102. Chen, C. et al. Quantized conductance of Majorana zero mode in the vortex of the topological superconductor (Li0. 84Fe0. 16) OHFeSe. *Chinese Phys. Lett.* **36**, 57403 (2019).
103. Law, K. T., Lee, P. A. & Ng, T. K. Majorana Fermion Induced Resonant Andreev Reflection. *Phys. Rev. Lett.* **103**, 237001 (2009).
104. Ruby, M. et al. Tunneling Processes into Localized Subgap States in Superconductors.

*Phys. Rev. Lett.* **115**, 87001 (2015).
105. Scheer, E., Joyez, P., Esteve, D., Urbina, C. & Devoret, M. H. Conduction Channel Transmissions of Atomic-Size Aluminum Contacts. *Phys. Rev. Lett.* **78**, 3535–3538 (1997).
106. Villas, A. *et al.* Interplay between Yu-Shiba-Rusinov states and multiple Andreev reflections. *Phys. Rev. B* **101**, 235445 (2020).
107. Chen, C. *et al.* Atomic line defects and zero-energy end states in monolayer Fe(Te,Se) high-temperature superconductors. *Nat. Phys.* **16**, 536–540 (2020).
108. Kane, C. L. & Mele, E. J. Quantum Spin Hall Effect in Graphene. *Phys. Rev. Lett.* **95**, 226801 (2005).
109. König, M. *et al.* Quantum Spin Hall Insulator State in HgTe Quantum Wells. *Science (80-. ).* **318**, 766 LP – 770 (2007).
110. Nilsson, J., Akhmerov, A. R. & Beenakker, C. W. J. Splitting of a Cooper Pair by a Pair of Majorana Bound States. *Phys. Rev. Lett.* **101**, 120403 (2008).
111. Mi, S., Pikulin, D. I., Wimmer, M. & Beenakker, C. W. J. Proposal for the detection and braiding of Majorana fermions in a quantum spin Hall insulator. *Phys. Rev. B* **87**, 241405 (2013).
112. Hart, S. *et al.* Induced superconductivity in the quantum spin Hall edge. *Nat. Phys.* **10**, 638–643 (2014).
113. Pribiag, V. S. *et al.* Edge-mode superconductivity in a two-dimensional topological insulator. *Nat. Nanotechnol.* **10**, 593–597 (2015).
114. Murani, A. *et al.* Ballistic edge states in Bismuth nanowires revealed by SQUID interferometry. *Nat. Commun.* **8**, 15941 (2017).
115. Lüpke, F. *et al.* Proximity-induced superconducting gap in the quantum spin Hall edge state of monolayer WTe2. *Nat. Phys.* **16**, 526–530 (2020).
116. Murakami, S. Quantum Spin Hall Effect and Enhanced Magnetic Response by Spin-Orbit Coupling. *Phys. Rev. Lett.* **97**, 236805 (2006).
117. Drozdov, I. K. *et al.* One-dimensional topological edge states of bismuth bilayers. *Nat. Phys.* **10**, 664–669 (2014).
118. Nayak, A. K. *et al.* Resolving the topological classification of bismuth with topological defects. *Sci. Adv.* **5**, eaax6996 (2019).
119. Jäck, B., Xie, Y., Andrei Bernevig, B. & Yazdani, A. Observation of backscattering induced by magnetism in a topological edge state. *Proc. Natl. Acad. Sci.* **117**, 16214 LP – 16218 (2020).
120. Sun, H.-H. *et al.* Coexistence of Topological Edge State and Superconductivity in Bismuth Ultrathin Film. *Nano Lett.* **17**, 3035–3039 (2017).
121. Tang, S. *et al.* Quantum spin Hall state in monolayer 1T'-WTe2. *Nat. Phys.* **13**, 683–687 (2017).
122. Shi, Y. *et al.* Imaging quantum spin Hall edges in monolayer WTe$_2$ *Sci. Adv.* **5**, eaat8799 (2019).
123. Huang, B. *et al.* Electrical control of 2D magnetism in bilayer CrI3. *Nat. Nanotechnol.* **13**, 544–548 (2018).
124. Wang, Q. *et al.* Large intrinsic anomalous Hall effect in half-metallic ferromagnet Co 3 Sn 2 S 2 with magnetic Weyl fermions. *Nat. Commun.* **9**, 1–8 (2018).
125. Fatemi, V. *et al.* Electrically tunable low-density superconductivity in a monolayer topological insulator. *Science (80-. ).* **362**, 926 LP – 929 (2018).

126. Chamon, C., Jackiw, R., Nishida, Y., Pi, S.-Y. & Santos, L. Quantizing Majorana fermions in a superconductor. *Phys. Rev. B* **81**, 224515 (2010).
127. Sticlet, D., Bena, C. & Simon, P. Spin and Majorana Polarization in Topological Superconducting Wires. *Phys. Rev. Lett.* **108**, 96802 (2012).
128. He, J. J., Ng, T. K., Lee, P. A. & Law, K. T. Selective Equal-Spin Andreev Reflections Induced by Majorana Fermions. *Phys. Rev. Lett.* **112**, 37001 (2014).
129. Haim, A., Berg, E., von Oppen, F. & Oreg, Y. Signatures of Majorana Zero Modes in Spin-Resolved Current Correlations. *Phys. Rev. Lett.* **114**, 166406 (2015).
130. Björnson, K., Pershoguba, S. S., Balatsky, A. V & Black-Schaffer, A. M. Spin-polarized edge currents and Majorana fermions in one- and two-dimensional topological superconductors. *Phys. Rev. B* **92**, 214501 (2015).
131. Kotetes, P., Mendler, D., Heimes, A. & Schön, G. Majorana fermion fingerprints in spin-polarised scanning tunnelling microscopy. *Phys. E Low-dimensional Syst. Nanostructures* **74**, 614–624 (2015).
132. Szumniak, P., Chevallier, D., Loss, D. & Klinovaja, J. Spin and charge signatures of topological superconductivity in Rashba nanowires. *Phys. Rev. B* **96**, 41401 (2017).
133. Wiesendanger, R. Spin mapping at the nanoscale and atomic scale. *Rev. Mod. Phys.* **81**, 1495–1550 (2009).
134. Cornils, L. *et al.* Spin-Resolved Spectroscopy of the Yu-Shiba-Rusinov States of Individual Atoms. *Phys. Rev. Lett.* **119**, 197002 (2017).
135. Jiang, K., Dai, X. & Wang, Z. Quantum Anomalous Vortex and Majorana Zero Mode in Iron-Based Superconductor Fe(Te,Se). *Phys. Rev. X* **9**, 11033 (2019).
136. Zhang, S. S. *et al.* Field-free platform for Majorana-like zero mode in superconductors with a topological surface state. *Phys. Rev. B* **101**, 100507 (2020).
137. Kot, P. *et al.* Microwave-assisted tunneling and interference effects in superconducting junctions under fast driving signals. *Phys. Rev. B* **101**, 134507 (2020).
138. González, S. A. *et al.* Photon-assisted resonant Andreev reflections: Yu-Shiba-Rusinov and Majorana states. *Phys. Rev. B* **102**, 45413 (2020).
139. Perrin, V., Civelli, M. & Simon, P. Discriminating Majorana from Shiba bound-states by tunneling shot-noise tomography. in (2020).
140. Naaman, O., Teizer, W. & Dynes, R. C. Fluctuation dominated Josephson tunneling with a scanning tunneling microscope. *Phys. Rev. Lett.* **87**, 97004 (2001).
141. Jäck, B. *et al.* Critical Josephson current in the dynamical Coulomb blockade regime. *Phys. Rev. B* **93**, 20504 (2016).
142. Randeria, M. T., Feldman, B. E., Drozdov, I. K. & Yazdani, A. Scanning Josephson spectroscopy on the atomic scale. *Phys. Rev. B* **93**, 161115 (2016).
143. Jiao, L. *et al.* Chiral superconductivity in heavy-fermion metal UTe2. *Nature* **579**, 523–527 (2020).
144. Wu, S. *et al.* Observation of the quantum spin Hall effect up to 100 kelvin in a monolayer crystal. *Science (80-. ).* **359**, 76 LP – 79 (2018).
145. Pikulin, D. Proposal for a scalable charging-energy-protected topological qubit in a quantum spin Hall system. in (2020).
146. Béri, B. & Cooper, N. R. Topological Kondo Effect with Majorana Fermions. *Phys. Rev. Lett.* **109**, 156803 (2012).
147. Madhavan, V., Chen, W., Jamneala, T., Crommie, M. F. & Wingreen, N. S. Tunneling into a Single Magnetic Atom: Spectroscopic Evidence of the Kondo Resonance. *Science*

(80-. ). **280**, 567 LP – 569 (1998).
148. Odobesko, A. B. *et al.* Preparation and electronic properties of clean superconducting Nb(110) surfaces. *Phys. Rev. B* **99**, 115437 (2019).
149. Yan, Z., Song, F. & Wang, Z. Majorana Corner Modes in a High-Temperature Platform. *Phys. Rev. Lett.* **121**, 96803 (2018).
150. Liu, T., He, J. J. & Nori, F. Majorana corner states in a two-dimensional magnetic topological insulator on a high-temperature superconductor. *Phys. Rev. B* **98**, 245413 (2018).
151. Hsu, Y.-T., Cole, W. S., Zhang, R.-X. & Sau, J. D. Inversion-Protected Higher-Order Topological Superconductivity in Monolayer ${\mathrm{WTe}}_{2}$. *Phys. Rev. Lett.* **125**, 97001 (2020).
152. Feldmeier, J., Natori, W., Knap, M. & Knolle, J. Local probes for charge-neutral edge states in two-dimensional quantum magnets. *Phys. Rev. B* **102**, 134423 (2020).
153. König, E. J., Randeria, M. T. & Jäck, B. Tunneling spectroscopy of quantum spin liquids. *Phys. Rev. Lett.* **125**, 267206 (2020).
154. Udagawa, M., Takayoshi, S. & Oka, T. STM as a single Majorana detector of Kitaev's chiral spin liquid. *arXiv Prepr. arXiv2008.07399* (2020).